\documentstyle[12pt,epsfig,a4]{article} 
\newcommand{\vs}{\vspace{-0.25cm}}
\begin{document} 

\begin{center}
{\Large{\bf Scales in nuclear matter: Chiral dynamics with pion nucleon form 
factors}}\footnote{Work supported in part by BMBF and GSI.}

\medskip

N. Kaiser, M. M\"uhlbauer and W. Weise\\

\smallskip

{\small Physik Department, Technische Universit\"{a}t M\"{u}nchen,
    D-85747 Garching, Germany}

\end{center}

\medskip

\begin{abstract}
A systematic calculation of nuclear matter is performed which includes the
long-range correlations between nucleons arising from one- and two-pion
exchange. Three-body effects from $2\pi$-exchange with excitations of virtual
$\Delta(1232)$-isobars are also taken into account in our diagrammatic
calculation of the energy per particle $\bar E(k_f)$. In order to eliminate
possible high-momentum components from the interactions we introduce at each 
pion-baryon vertex a form factor of monopole type. The empirical nuclear 
matter saturation point, $\rho_0 \simeq 0.16\,$fm$^{-3}$, $\bar E_0\simeq 
-16\,$MeV, is well reproduced with a monopole mass of $\Lambda \simeq 4\pi 
f_\pi \simeq 1.16\,$GeV. As in the recent approach based on the universal 
low-momentum $NN$-potential $V_{\rm low-k}$, the inclusion of three-body
effects is crucial in order to achieve saturation of nuclear matter. We
demonstrate that the dependence of the pion-exchange contributions to $\bar
E(k_f)$  on the ''resolution'' scale $\Lambda$ can be compensated over a wide 
range of $\Lambda$ by counterterms with two ''running'' contact-couplings. As 
a further application we study the in-medium chiral condensate $\langle \bar q 
q \rangle( \rho)$ beyond the linear density approximation. For $\rho \leq 1.5
\rho_0$ we find small corrections from the derivative $d \bar  E(k_f)/d m_\pi$,
which are stable  against variations of the monopole regulator mass $\Lambda$. 
\end{abstract}

\bigskip 

\bigskip

\section{Introduction and summary} 
In recent years a novel approach to the nuclear matter problem based on
effective field theory has emerged. Its key element is a separation of long- 
and short-distance dynamics and an ordering scheme in powers of small
momenta. At nuclear matter saturation density $\rho_0 \simeq 0.16\,$fm$^{-3}$ 
the Fermi momentum $k_{f0}$ and the pion mass $m_\pi$ are comparable scales 
($k_{f0}\simeq 2 m_\pi$). Therefore pions must be included as explicit 
degrees of freedom in the description of the nuclear many-body dynamics. The 
contributions to the energy per particle, $\bar E(k_f)$, of isospin-symmetric 
(spin-saturated) nuclear matter as they originate from chiral pion-nucleon 
dynamics have been computed up to three-loop order in refs.\cite{lutz,nucmat}. 
Both calculations are able to reproduce the empirical saturation 
point by adjusting one single parameter (either a contact-coupling $g_0+g_1 
\simeq 3.23$ \cite{lutz} or a momentum cut-off $\Lambda \simeq 0.65 \,$GeV 
\cite{nucmat}) related to unresolved short-distance dynamics. The mechanism 
for saturation in these  approaches is mainly a repulsive contribution to the 
energy per particle generated by Pauli-blocking in second order (iterated) 
one-pion exchange. As outlined in ref.\cite{nucmat} this mechanism becomes 
particularly transparent by taking the chiral limit $m_\pi= 0$. In that case 
the interaction contributions to the energy per particle are completely 
summarized by an attractive $k_f^3$-term and a repulsive $k_f^4$-term where
the parameter-free prediction for the coefficient of the latter is very close
to the one extracted from a realistic nuclear matter equation of state.   

This chiral approach to nuclear matter has been extended and improved in 
ref.\cite{deltamat} by including systematically the effects from
$2\pi$-exchange with excitation of virtual $\Delta(1232)$-isobars. The physical
motivation for such an extension is threefold. First, the spin-isospin-$3/2$
$\Delta(1232)$-resonance is the most prominent feature of low-energy $\pi
N$-scattering. Secondly, it is well known that the $2\pi$-exchange between
nucleons with excitation of virtual $\Delta$-isobars generates the medium- and
long-range components of the isoscalar central $NN$-attraction \cite{2pidel}.
In phenomenological one-boson exchange models this $\pi N\Delta$-dynamics is 
often simulated by a scalar ''$\sigma$''-meson exchange. Thirdly, the 
delta-nucleon mass-splitting $\Delta = 293\,$MeV is of the same size as the 
Fermi momentum $k_{f0} \simeq 2m_\pi$ at nuclear matter saturation density. 
Therefore pions and $\Delta$-isobars should both be treated as explicit
degrees of freedom.  

It has been found in ref.\cite{deltamat} that the inclusion of the chiral $\pi 
N\Delta$-dynamics significantly improves e.g. the momentum-dependence of the 
(real) single-particle potential $U(p,k_f)$ and the isospin properties (as
revealed in the density-dependent asymmetry energy $A(k_f)$ and the neutron 
matter equation of state $\bar E_n(k_n)$). However, there remain some open 
questions in these perturbative calculations of nuclear matter. The effective
short-range terms are adjusted directly to nuclear matter bulk properties and 
thus seem unrelated to those relevant for free $NN$-scattering. Also, no 
deeper justification for a perturbative treatment of nuclear matter (besides
of being very successful) could be given. 

Important progress in this direction has come recently from the work of Bogner
et al.\,\cite{achim} based on the universal low-momentum $NN$-potential $V_{\rm
low-k}$. This potential operates by construction only between the low-momentum 
nucleon states, $|\vec p\,|\leq 0.4\,$GeV, where it is truly determined by 
elastic $NN$-scattering data. It has been demonstrated in ref.\cite{achim}
that, due to the absence of a model-dependent short-range repulsive core in
the potential $V_{\rm low-k}$, its second and higher order iterations in
nuclear matter (i.e. the successive terms in the Brueckner ladder-series) turn
out to be small. The primary reason for this (unconventional) feature is the
Pauli exclusion principle: with increasing Fermi momentum $k_f$ Pauli-blocking
reduces progressively the available low-momentum phase space wherein $V_{\rm 
low-k}$ acts. Under such conditions a perturbative treatment of the two-body 
interaction in  nuclear matter turns out to be justified. It has also been 
demonstrated in ref.\cite{achim} that a saturation of nuclear matter cannot be
achieved from the (low-momentum) two-body interaction alone (see also
ref.\cite{kukei} for earlier Hartree-Fock calculations with $V_{\rm low-k}$ 
exhibiting the same feature). Repulsive three-nucleon interactions (in
particular the long-range ones induced by $2\pi$-exchange) are crucial in
order to stabilize the nuclear many-body system against collapse. As
emphasized in ref.\cite{achim} in an effective low-energy theory (with a 
''spatial resolution'' of $\lambda=(0.4\,{\rm GeV})^{-1}\simeq 0.5\,$fm as it 
is inherent to the $V_{\rm low-k}$ potential) many-body forces are inevitable. 
Taking into account furthermore the dominant second order contributions from  
$V_{\rm low-k}$ and the three-nucleon force (adjusted partly to triton and 
$^4$He binding energies) a saturation minimum of $\bar E_0 \simeq -12 \,$MeV 
at the correct equilibrium density $k_{f0} \simeq 1.35 \,$fm$^{-1}$ has been 
obtained \cite{achim}. Presently, there is ongoing activity to further improve 
these calculations in order to close the remaining  small gap to the empirical 
binding energy per particle, $\bar E_0 \simeq -16 \,$MeV.

The purpose of the present paper is to reanalyze the chiral approach to 
nuclear matter from this new perspective and to establish connections 
to the $V_{\rm low-k}$ approach (which actually starts from phase-shift 
equivalent $NN$-potentials). As already mentioned, the second-order tensor 
force from iterated one-pion exchange plays an essential role in order to 
obtain binding and saturation of nuclear matter in the (perturbative) chiral 
approaches studied so far \cite{lutz,nucmat,deltamat}. On the other hand its 
strong short-distance components are one of the sources which render the 
Brueckner ladder-series in nuclear matter non-perturbative (i.e. not 
convergent) \cite{achim}. It is therefore advisable to eliminate the
high-momentum components from the pion-induced $NN$-interactions. For that 
purpose we follow the phenomenology of the one-boson exchange potentials and 
introduce at each pion-nucleon interaction vertex a ''form factor'' of 
monopole type:
\begin{equation} 
F(q^2)= {\Lambda^2-m_\pi^2 \over \Lambda^2-q^2} \,,
\end{equation}       
with $\Lambda $ the monopole mass. Here, $q^2\leq 0$ denotes the squared 
four-momentum transfer carried by the virtual pion. Typical values of the
monopole mass $\Lambda$ employed in one-boson exchange potentials lie in the 
range: $1.0\,{\rm GeV} < \Lambda < 1.7\,$GeV \cite{mach,cdbonn}. With the form 
factor $F(q^2)$ included as a regulator on high-momentum components the 
$1\pi$- and $2\pi$-exchange contributions to the energy per particle $\bar
E(k_f)$ depend on the monopole mass $\Lambda$. As a first orientation we find 
that the ''natural choice'' $\Lambda \simeq 4\pi f_\pi \simeq 1.16\,$GeV (the 
chiral symmetry breaking scale) reproduces correctly the empirical nuclear 
matter saturation  point: $\rho_0 = 2k_{f0}^3/3\pi^2 \simeq 0.16\,$fm$^{-3}$, 
$\bar E(k_{f0}) \simeq -16 \,$MeV. As in the $V_{\rm low-k}$ approach of 
ref.\cite{achim} the inclusion of the pion-induced three-body terms (with and 
without virtual $\Delta$-isobar excitation) is essential in order to achieve 
saturation of nuclear matter. 

However, a physical low-energy quantity such as the nuclear 
matter equation of state $\bar E(k_f)$ should not depend on a parameter 
$\Lambda$ which merely scans the ''spatial resolution'' of the pion-baryon 
interactions involved.  We demonstrate that this (unphysical) $\Lambda
$-dependence of the pion-exchange contributions is, over a large range of 
monopole masses and densities, almost perfectly counterbalanced by the $k_f^3$-
and $k_f^5$-terms related to two ''running'' short-distance contact-couplings 
$B_{3,5}(\Lambda)$. 

Having reached a description of nuclear matter in terms of explicit 
pion-exchange dynamics, which is stable against variations of the
short-distance scale $\Lambda$,  we can study the in-medium chiral condensate 
$\langle \bar q q\rangle(\rho)$ beyond the linear density approximation. These 
corrections are obtained by differentiating the calculated energy density of 
nuclear matter $\rho \bar E(k_f)$ with respect to the pion mass (or 
equivalently, the light quark mass $m_q$). Below the saturation density $\rho 
\leq 0.16\,$fm$^{-3}$ we find very small corrections. At higher densities a 
tendency counteracting chiral restoration sets in. Moreover, there is little 
dependence of the derivative $d \bar E(k_f)/d m_\pi$ on the monopole mass 
$\Lambda$ (assuming the short-distance  contact-couplings $B_{3,5}(\Lambda )$ 
to be  independent of the quark mass).      
                       
Our paper is organized as follows: In section 2 we present first the analytical
expressions for the one- and two-pion exchange contributions to the energy per
particle $\bar E(k_f)$ including the monopole form factors. Then we discuss 
in section 3 the results for the nuclear matter equation of state together
with the compensating short-distance contact-couplings $B_{3,5}(\Lambda)$.
Section 4 is devoted to the in-medium chiral quark condensate $\langle \bar q q
\rangle (\rho)$. Section 5 ends finally with some concluding remarks and an
outlook.   

\begin{figure}
\begin{center}
\includegraphics[scale=0.9]{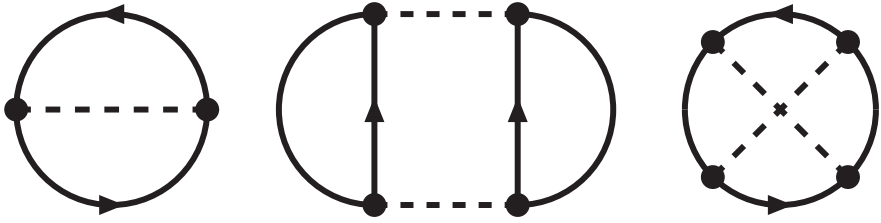}
\end{center}
\vspace{-0.5cm}
\caption{Two-loop one-pion exchange Fock diagram and three-loop iterated 
one-pion exchange Hartree and Fock diagrams. The combinatoric factors are
$1/2$, $1/4$ and $1/4$, in the order shown. Their isospin factors for 
symmetric nuclear matter are $6$, $12$ and $-6$, respectively.}
\end{figure}

\section{Diagrammatic calculation of the energy per particle}
The first contribution to the energy per particle $\bar E(k_f)$ is the kinetic
energy of relativistic Fermi gas of nucleons (expanded in powers of $1/M_N$):
\begin{equation}
\bar E(k_f)^{(\rm kin)}= {3k_f^2\over 10 M_N}-{3k_f^4\over 56 M_N^3}+{k_f^6 
\over 48 M_N^5}\,. \end{equation}
Here, $k_f$ denotes the Fermi momentum, related to the nucleon density in the
usual way, $\rho = 2k_f^3/3\pi^2$, and $M_N=939\,$MeV stands for the (average)
nucleon mass. At the densities of interest, $\rho \leq 0.5\,$fm$^{-3}$, the 
last term in eq.(2) is already negligibly small ($<0.1\,$MeV). Next, we
enumerate the interaction contributions from one- and two-pion exchange. For
each diagram we present only the final result omitting all technical details 
related to solving the loop and Fermi sphere integrals in the presence of the
monopole form factors. 
\subsection{One-pion exchange Fock term}
The contribution of the left $1\pi$-exchange Fock diagram in Fig.\,1 
(including the relativistic $1/M_N^2$-correction) reads:
\begin{eqnarray}
\bar E(k_f)^{(1\pi)} &=&{3 g_A^2 m_\pi^3 \over (4\pi f_\pi)^2 }\, \Bigg\{
{r^4-3r^2+2\over 8r^4u}+\arctan(2u)+{1-3r^2\over 2 r^3}\arctan(2ru) 
\nonumber\\ & & -{1+12u^2\over 32u^3}\ln(1+4u^2)+ {3r^2-2+12 r^2u^2(2r^2-1) 
\over 32r^6u^3}\ln(1+4r^2u^2)\nonumber\\ & &+{m_\pi^2\over 40 M_N^2} 
\bigg[{r^6-4r^2+3\over 2r^6u}+{3u\over r^4}(3r^4-7r^2+4)-(5+12u^2) 
\arctan(2u)\nonumber \\ && +{1\over 2r^5}(25r^2-15-12r^2u^2+36r^4 u^2)\arctan
(2ru) -{1\over 8u^3}\ln(1+4u^2)\nonumber \\ &&+{4r^2-3\over 8r^8u^3}\ln(1+
4r^2u^2)\bigg] \Bigg\}\,. \end{eqnarray} 
For notational simplicity we have introduced the dimensionless variable 
$u =k_f /m_\pi$ and the ratio $r= m_\pi/\Lambda$. The other occurring 
parameters are: $g_A = 1.3$ (the nucleon axial-vector coupling constant),
$m_\pi = 135\, $MeV (the neutral pion mass), and $f_\pi = 92.4\,$MeV (the weak
pion-decay constant). In the limit $r\to 0$ one recovers the result eq.(6) in
ref.\cite{nucmat} without the monopole form factor. 
\subsection{Iterated one-pion exchange}
The Hartree diagram of iterated $1\pi$-exchange (middle diagram in Fig.\,1)
with two medium insertions\footnote{Medium insertion is technical notation for
the difference between the nucleon propagator in the medium and in the vacuum
(see sect.\,2 in ref.\cite{nucmat}). To each medium insertion belongs an
integration over a  Fermi sphere of radius $k_f$.} leads to the following
analytical expression:  
\begin{eqnarray} 
\bar E(k_f)^{\rm (H2)}&= & {3 g_A^4M_N m_\pi^4\over 5 (8\pi)^3 f_\pi^4} \Bigg\{
{20(3+r^2)+32(1+r^2)u^2\over 1-r^2}\Big[ \arctan(2u)-\arctan(2r u)\Big]
\nonumber \\ & & + {1 \over u}\bigg( {1\over 2}+{5\over 16r^5}-{11\over
8r^3}+{41 \over 16 r}+{4 \over 1+r} \bigg) + u \bigg( 29-{15\over 8r^3}+{41 
\over 4r} -{75r  \over 8} -{88 \over 1+r} \bigg) \nonumber \\ & &  +{r^2-9-20(7
+r^2)u^2 \over 8(1-r^2)u^3}\,\ln(1+4u^2)\\ & & +{27 r^2-5-63r^4+ 105r^6+20r^2
u^2(1-7r^2+35r^4+35r^6)\over 64r^7(1-r^2)u^3}\ln(1+4r^2u^2)  \Bigg\}    \,. 
\nonumber \end{eqnarray}
The corresponding exchange term comes from the Fock diagram of iterated
$1\pi$-exchange (right diagram in Fig.\,1) with two medium insertions. While
the loop integral including the two squared monopole form factors can be 
solved there remains a single non-elementary integral from the integration
over the Fermi spheres. We get:
\begin{eqnarray}
\bar E(k_f)^{\rm (F2)} &= &{3 g_A^4M_N m_\pi^4 \over(4\pi u )^3 f_\pi^4}\, \!
\int_0^u\! dx \, x(u-x)^2 (2u+x) \Bigg\{ {1+8x^2+8x^4\over 2+4x^2}\arctan x
\nonumber \\ & & -{(1-r^2)^2(1+4x^2)\arctan(2x)\over(2+4x^2)(1+r^2+4r^2x^2)^2}
+\bigg[1+2x^2-{1+r^4+4r^4x^2\over(1+r^2+4r^2x^2)^2}\bigg]\nonumber\\ && \times
\bigg[\arctan{4r^2x\over1-r^2+4r^2x^2}-\arctan{2rx\over 1+r}-\arctan{2rx\over 
1-r}\bigg]\nonumber \\& & +\bigg[1+2x^2 - {1+r^4+4r^4 x^2(1+r^2+2r^4x^2) \over 
4r^2(1+2r^2x^2)^3} \bigg] \arctan(rx) \nonumber\\& & +\bigg[{1+r^4+4r^4x^2(1+ 
r^2+2r^4x^2) \over 4r^2(1+2r^2x^2)^3} -{1+r^4+4r^4x^2 \over (1+r^2+4r^2x^2)^2} 
\bigg] \arctan(2rx) \nonumber\\  & & +x(1-r^2)\bigg[{1+r^{-1}\over 1+r^2+4r^2
x^2}+{(1-r)^2+4r^2x^2\over (1+r^2+4r^2x^2)^2-4r^2}\nonumber\\& & + {4rx^2  
(r^2-2)+5r-9r^{-1}\over 16 (1+r^2x^2)^2}-{r^{-1}+r x^2(1+r^2)
\over (1+2r^2x^2)^2}\bigg]\Bigg\}\,. \end{eqnarray}
In our way of organizing the many-body calculation, the Pauli-blocking effects 
are represented by diagrams with three medium insertions 
\cite{nucmat,deltamat}. The contribution of the Hartree diagram with three
medium insertions (including the fourth power of a monopole form factor)
reads:  
\begin{eqnarray}
\bar E(k_f)^{(\rm H3)} & = & {9 g_A^4 M_N m_\pi^4 \over (4\pi f_\pi)^4 u^3}
\!\int_0^u\! dx\,x^2\! \int_{-1}^1\! dy \,\bigg[ 2uxy+(u^2-x^2 y^2) \ln{u+xy 
\over u-xy} \bigg]\nonumber \\& & \times  \bigg\{{2+2r^2\over 1-r^2}\Big[\ln(1
+r^2s^2)-\ln(1+s^2)\Big]+{2s^2+s^4 \over 1+s^2} \nonumber \\ && + { r^2s^2 
\over 3(1+r^2s^2)^3} \Big[6-6s^2+15r^2s^2-8r^2s^4 +7 r^4s^4+r^6s^4-3r^4s^6 
\Big]\bigg\}\,, \end{eqnarray}
with the abbreviation  $s = xy +\sqrt{u^2-x^2+x^2y^2}$. On the other hand one
gets from the right Fock diagram in Fig.\,1 with three medium insertions
(including two squared monopole form factors): 
\begin{eqnarray}
\bar E(k_f)^{(\rm F3)} & = & {9 g_A^4 M_N m_\pi^4 \over (4\pi f_\pi)^4 u^3}
\int_0^u\! dx \Bigg\{{G^2\over 8}-{x^2\over 4}\!\int_{-1}^1\! dy\,\!\int_{
-1}^1\! dz\,{yz\, \theta(y^2+z^2-1)\over |yz|\sqrt{y^2+z^2-1}} \bigg[{(1-r^2)
s^2\over 1+r^2s^2} \nonumber \\ & & -\ln(1+s^2) +\ln(1+r^2s^2)\bigg]
\bigg[{(1-r^2)t^2\over 1+r^2t^2}-\ln(1+t^2)+\ln(1+r^2t^2)\bigg]\Bigg\}\,,
\end{eqnarray}
with another abbreviation  $t = xz +\sqrt{u^2-x^2+x^2z^2}$ and the auxiliary
function:  
\begin{eqnarray}
G &=&{u \over r^2}(1-r^2) -{1\over 4 x} [1+(u+x)^2] [1+(u-x)^2] \ln{1+(u+x)^2
\over 1+(u-x)^2}\nonumber \\ && +{1\over 4x} \Big[ 2r^{-2}
-r^{-4}+(u^2-x^2)^2+2u^2+2x^2 \Big] \ln{1+r^2(u+x)^2\over 1+r^2(u-x)^2}\,.
\end{eqnarray}
Note that the contributions in eqs.(4--7) carry the large scale enhancement
factor $M_N$. It stems from the energy denominator of these iterated diagrams
which is proportional to the difference of small nucleon kinetic energies. 
\subsection{Irreducible two-pion exchange: Spectral representation} 
Next, we come to the irreducible $2\pi$-exchange contributions (with no,
single, and double $\Delta(1232)$-isobar excitation). Their direct evaluation 
requires the one-loop $NN$-scattering T-matrix with a monopole form factor
attached to each pion-baryon vertex. The essential long-distance information 
about these $2\pi$-exchange one-loop diagrams is already contained in the 
spectral functions (or imaginary parts). The effect of the monopole form
factors\footnote{We choose the same monopole form factor for each pion-baryon
vertex.} on the spectral functions can be easily inferred from the following 
partial fraction decomposition: 
\begin{equation} {1\over m_\pi^2-q^2}\,\bigg({\Lambda^2-m_\pi^2 \over \Lambda^2
-q^2}\bigg)^2 = {1\over m_\pi^2-q^2} -{1\over \Lambda^2 -q^2}- {\Lambda^2-
m_\pi^2 \over (\Lambda^2-q^2)^2}\,. \end{equation} 
It instructs us that a pion-exchange with monopole form factors attached to
the vertices is identical to a point-like pion-exchange minus additional  
exchanges involving a ''heavy particle'' of mass $\Lambda$. By unitarity this
implies that the spectral function of a one-loop $2\pi$-exchange diagram with 
monopole form factors has thresholds at $\mu = 2m_\pi$, $\mu =\Lambda+m_\pi$ 
and $\mu =2\Lambda$. At the higher thresholds the effective chiral Lagrangian 
is, however, no more applicable. In this case a meaningful and physically 
reasonable way to include effects from the monopole form factor is to cut off
the spectral integral at $\mu= \Lambda +m_\pi$. Putting all pieces together
\cite{deltamat} this procedure leads to the following $2\pi$-exchange two-body
Fock term:            
\begin{eqnarray} 
\bar E(k_f)^{(2\pi)} & = & {1 \over 8\pi^3} \int_{2m_\pi}^{\Lambda+m_\pi}
\!\! d\mu\,{\rm Im}(V_C+3W_C+2\mu^2V_T+6\mu^2W_T) \bigg\{ 3\mu k_f -{4k_f^3
\over 3\mu }\nonumber \\ & &+{8k_f^5\over 5\mu^3}-{\mu^3 \over 2k_f}-4\mu^2
\arctan{2k_f\over\mu} +{\mu^3 \over 8k_f^3}(12k_f^2+\mu^2) \ln\bigg( 1+{4k_f^2
  \over \mu^2} \bigg) \bigg\}\,, \end{eqnarray}
where Im$V_C$, Im$W_C$, Im$V_T$ and Im$W_T$ are the spectral functions of the
isoscalar and isovector central and tensor $NN$-amplitudes, respectively. 
Explicit expressions of these imaginary parts for the contributions of the 
triangle diagram with single $\Delta$-excitation and the box diagrams with 
single and double $\Delta$-excitation can be easily constructed from the 
analytical formulas given in section 3 of ref.\cite{2pidel}. Note that the 
$\mu$- and $k_f$-dependent weighting function in eq.(10) involves two 
subtractions. This way it is guaranteed that the spectral integral at low and 
moderate densities is dominated by low invariant $\pi\pi$-masses, $2m_\pi< \mu 
<1\,$GeV. 

The contributions to the energy per particle from irreducible 
$2\pi$-exchange (with only nucleon intermediate states) can also be cast into 
the form eq.(10). The corresponding non-vanishing spectral functions read: 
\begin{equation} {\rm Im}W_C = {\sqrt{\mu^2-4m_\pi^2} \over 3\pi 
\mu (4f_\pi)^4} \bigg[ 4m_\pi^2(1+4g_A^2-5g_A^4) +\mu^2(23g_A^4-10g_A^2-1) + 
{48 g_A^4 m_\pi^4 \over \mu^2-4m_\pi^2} \bigg] \,, \end{equation}
\begin{equation} {\rm Im}V_T = - {6 g_A^4 \sqrt{\mu^2-4m_\pi^2} \over 
\pi  \mu (4f_\pi)^4}\,. \end{equation}
\subsection{Three-body terms with intermediate $\Delta(1232)$-isobars}
Finally, we come to the additional three-body terms which arise from
Pauli-blocking of intermediate nucleon states. The corresponding closed
Hartree and Fock $2\pi$-exchange diagrams with single virtual $\Delta$-isobar
excitation are shown in Fig.\,2. Their isospin factors are $8$, $0$ and $8$,
in the order shown. For the (left) three-loop Hartree diagram the integral 
over the product of  three Fermi spheres of radius $k_f$ can be solved in 
presence of the monopole form factors and its contribution to the energy per 
particle reads:

\begin{figure}
\begin{center}
\includegraphics[scale=0.9]{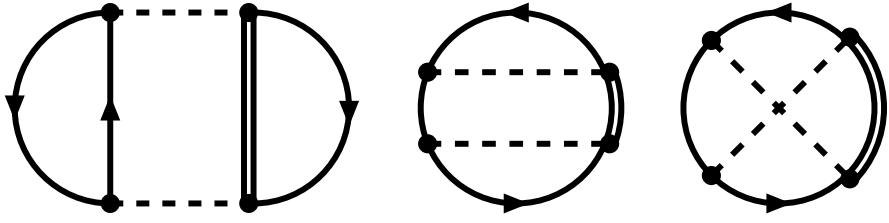}
\end{center}
\vspace{-0.5cm}
\caption{Hartree and Fock three-body diagrams related to $2\pi$-exchange with 
single virtual $\Delta$-isobar excitation. For symmetric nuclear matter the 
isospin factors are $8$, $0$ and $8$, in the order shown. The combinatoric
factor is $1$ for each diagram.} 
\end{figure}

\begin{eqnarray}
\bar E(k_f)^{(\rm H3,\Delta)}& = & {g_A^4 m_\pi^6 \over \Delta(2\pi f_\pi)^4}
\Bigg\{{1+9u^2+3r^2u^2\over 4(1-r^2)}\Big[\ln(1+4r^2u^2)-\ln(1+4u^2)\Big] 
\nonumber\\ & & +{4r^2u^2+u^4(2r^2-1+15r^4)\over 4r^2(1+4r^2u^2)}+{(5+3r^2)u^3
\over 1- r^2}\arctan(2u) \nonumber\\& & +{(1-15r^2-45r^4-5r^6)u^3 \over 8r^3
(1-r^2)} \arctan(2r u) \Bigg\}\,, \end{eqnarray}
with $u = k_f/m_\pi$ and $r = m_\pi/\Lambda$. The nonrelativistic $\Delta
$-propagator shows up in this expression merely via the (reciprocal)
mass-splitting $\Delta = 293\,$MeV. Furthermore, we have inserted in eq.(13) 
already the empirically well-satisfied relation $g_{\pi N \Delta} = 3g_{\pi 
N}/\sqrt{2}$ together with the Goldberger-Treiman relation $ g_{\pi N}= g_A 
M_N/f_\pi=13.2$. Finally, the contribution of the right three-body Fock 
diagram in Fig.\,2, with two squared monopole form factors included, is given 
by:  
\begin{equation} 
\bar E(k_f)^{(\rm F3,\Delta)}= -{3 g_A^4 m_\pi^6 \over 4\Delta(4\pi f_\pi)^4 
u^3} \int_0^u\,dx \Big(2 G^2_S+G^2_T \Big) \,,  \end{equation}
where we have introduced the two auxiliary functions:
\begin{eqnarray}
G_S &=& 4x \,\big[\arctan(u+x)+\arctan(u-x)\big]+(x^2-u^2-1)\ln{1+(u+x)^2\over 
1+(u-x)^2}\nonumber \\ & &+{2x\over r^3}(1-3r^2) \big\{\arctan[r(u+x)]+ 
\arctan[r(u-x)]\big\} \nonumber \\ & &+(u^2-x^2+2r^{-2}-r^{-4}) \ln{1+r^2(u+x
)^2\over 1+r^2(u-x)^2}\,, \end{eqnarray} 
\begin{eqnarray}
G_T & = & {1+u^2-x^2\over 8x^2}\,[1+(u+x)^2]\,[1+(u-x)^2]\,\ln{1+(u+x)^2\over 
1+(u-x)^2}\nonumber\\ & &+{(1-r^2)u\over 2r^4 x}\big[r^2(1-u^2+x^2)-2 \big]
+ {1\over 8}\bigg[ x^4+(1-3u^2)x^2 +2u^2+3u^4 +r^{-4} \nonumber \\ &&-2r^{-2} 
-{u^4 \over x^2}(3+u^2) + {3u^2 \over r^4 x^2}(1-2r^2) + {2-3r^2\over r^6 x^2} 
\bigg] \ln{1+r^2(u+x)^2\over1+r^2(u-x)^2}\,. \end{eqnarray}
Evidently, the three-body Fock term in eq.(14) is attractive and reduces, by 
about 1/3, the stronger  repulsive Hartree term in eq.(13).  

\section{Results for the nuclear matter equation of state}
We are now in the position to present numerical results for the nuclear matter
equation of state. The interaction contributions to the energy per particle 
$\bar E(k_f)$ written down in section 2 depend on a free parameter, the
monopole mass $\Lambda$. A first possible option is to adjust it to (one 
coordinate of) the empirical saturation point. Imposing a minimum of the
saturation curve $\bar E(k_f)$ at $k_{f0} = 262\,$MeV (corresponding to an
equilibrium density of $\rho_0 = 0.158\,$fm$^{-3}$) fixes the monopole mass to
$\Lambda_0=1.145\,$GeV. This value is surprisingly close to the chiral 
symmetry breaking scale  $\Lambda_\chi=4\pi f_\pi \simeq 1.16\,$GeV. It is 
also compatible with the typical monopole masses employed in one-boson 
exchange potentials \cite{mach}. With that fixed $\Lambda_0$-value the energy 
per particle at the saturation minimum comes out as $\bar E_0 =-16.7\, $MeV, 
in good agreement with the empirical value $\bar E_0=(-16 \pm 1)\,$MeV. 
This quite a non-trivial result. 

The decomposition of  $\bar E_0$ into contribution from the kinetic energy,
$1\pi$- and $2\pi$-exchange, as well as two- and three-body terms is also 
interesting: $\bar E_0= (21.64+ 13.63 -72.86-6.41+ 19.06-1.46+4.63+7.72-2.67)
\,$MeV, where the nine entries correspond to the terms in eqs.(2-7,10,13,14),
in that order. The full line in Fig.\,3 shows the resulting nuclear matter
equation of state as a function of the nucleon density $\rho = 2k_f^3/3\pi^2$
up to about $3\rho_0 \simeq 0.5\,$fm$^{-3}$. The curvature at its minimum
translates into a nuclear matter compressibility of $K = k_{f0}^2 \bar
E''(k_{f0}) = 292\,$MeV. It is about $10\%$ larger than the recent
extrapolation from giant monopole resonances of heavy nuclei, which gave the
value $K = (260 \pm 10)\, $MeV 
\cite{dario}.    

\begin{figure}
\begin{center}
\includegraphics[scale=0.5]{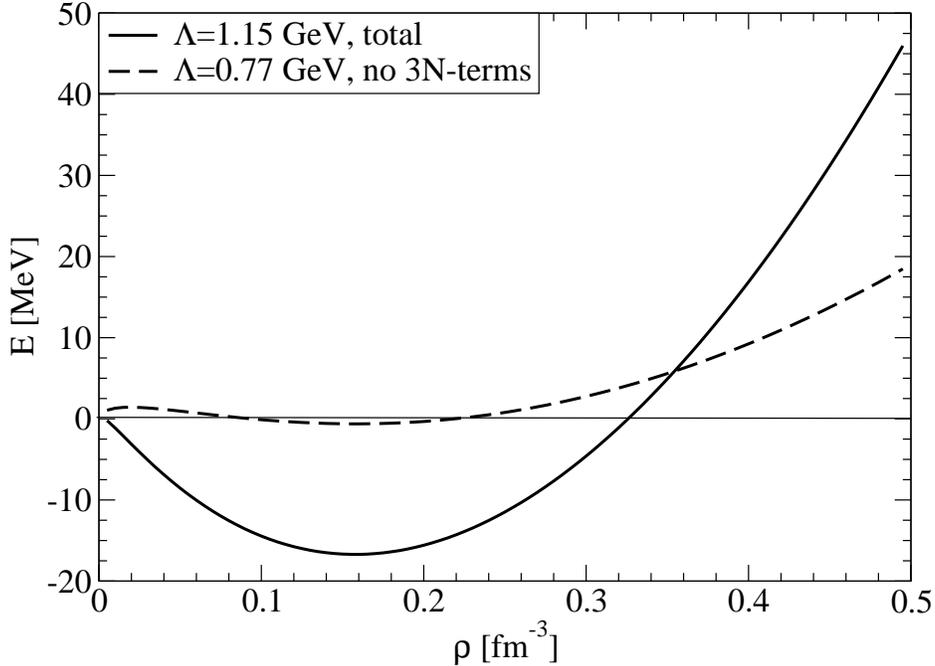}
\end{center}
\vspace{-0.5cm}
\caption{Energy per particle $\bar E(k_f)$ of isospin-symmetric nuclear matter
versus the nucleon density $\rho = 2k_f^3/3\pi^2$. In the dashed line all
three-body terms have been omitted.}  
\end{figure}

The inclusion of the $2\pi$-exchange three-body terms eqs.(6,7,13,14) is
crucial in order to achieve realistic saturation of nuclear matter in our
calculation. This feature is in accordance with the findings in the work of 
ref.\cite{achim} based on the universal low-momentum $NN$-potential $V_{\rm
low-k}$. It is also compatible with the results of sophisticated many-body 
calculations by the Urbana group \cite{akmal}. The dashed line in Fig.\,3 
shows the effect of turning off the $2\pi$-exchange three-body terms 
eqs.(6,7,13,14) in our calculation. After such a truncation there is almost no 
trace of nuclear matter saturation left over. In order to keep still a very 
shallow minimum at $k_{f0}=262\,$MeV the monopole mass has been reduced to 
$\Lambda = 0.77\,$GeV. Leaving $\Lambda_0 = 1.15\,$GeV unchanged the curve for 
$\bar E(k_f)$ without three-body terms would just decrease monotonically for 
all densities $0< \rho < 0.5\, $fm$^{-3}$, reaching the huge negative value 
$-85\,$MeV at its lower end.      

The pion-nucleon monopole form factor $F(q^2)$ introduced in our calculation
is only a regulator eliminating high momentum components. It is not a physical 
(observable) quantity. Interpreted differently, the monopole mass $\Lambda$ is 
a control parameter which monitors the ''spatial resolution'' at the 
pion-baryon interaction vertices. A low-energy quantity, such as the nuclear 
matter equation of state, should not depend it. We take the point of view,
characteristic of an effective field theory, that the explicit pion-exchange 
terms are accompanied by additional ''unresolved'' short-distance 
contributions. The most general contact-interaction of four nucleons 
(momentum-independent and quadratic in momenta) gives rise to the  following 
counterterm contribution to the energy per particle: 
\begin{equation} \bar E(k_f)^{(\rm ct)} = B_3(\Lambda) \, {k_f^3 \over M_N^2} +
  B_5(\Lambda) \, {k_f^5 \over M_N^4} \,, \end{equation} 
with $B_3(\Lambda)$ and $B_5(\Lambda)$ two dimensionless coupling
strengths. In the sense of a ''renormalization group equation'' we allow these
contact-couplings to run with the resolution scale $\Lambda$ such that the
physical observable, the total energy per particle $\bar E(k_f)$, is (as 
closely as possible) independent of $\Lambda$.        

\begin{figure}
\begin{center}
\includegraphics[scale=0.5]{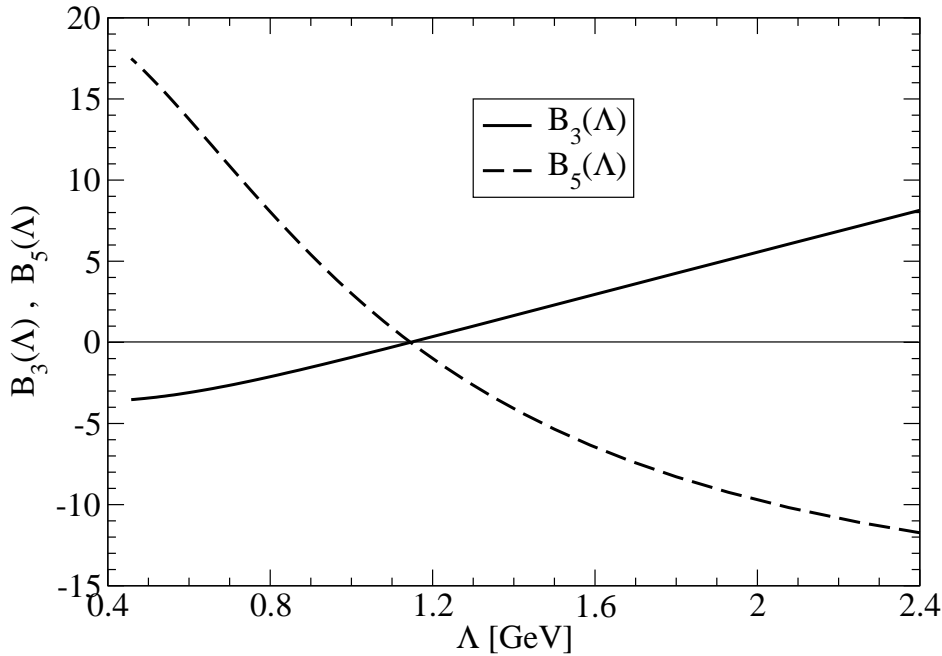}
\end{center}
\vspace{-0.3cm}
\caption{The ''running'' contact-couplings $B_3(\Lambda)$ and $B_5(\Lambda)$
as a function of the resolution scale $\Lambda$.}
\end{figure}

Fig.\,4 shows how the two contact-couplings $B_3(\Lambda)$ and $B_5(\Lambda)$
must vary with the monopole mass $\Lambda$ if the nuclear matter saturation 
point is required to stay fixed. The approximate linear rise of $B_3(\Lambda)$ 
over the whole range $0.4\,{\rm GeV}<\Lambda <2.4\,$GeV finds its explanation 
in the fact that for the iterated $1\pi$-exchange contributions eqs.(3,5), the 
monopole form factor serves also as a regulator on a linearly divergent loop 
integral (scaling as $1/r \sim \Lambda$) which contributes to the energy per 
particle linear in density ($\rho \sim k_f^3$).  

In Fig.\,5 we show the resulting nuclear matter equation of state for three 
different monopole masses $\Lambda =0.5\,$GeV, $1.0\,$GeV and  $1.5\,$GeV. In
the  density region $0 <\rho < 0.4\,$fm$^{-3}$ there is almost no dependence 
on the resolution scale $\Lambda$ left over. As expected, the small spreading 
of the three curves increases with increasing Fermi momentum $k_f$, where 
higher momentum components of the interactions get probed more sensitively. It 
is also astonishing to observe that the rather complicated density dependence 
of the $r$-dependent terms ($r=m_\pi/\Lambda$) written down in section 2 can 
be almost perfectly counterbalanced by two leading powers of the Fermi
momentum, $k_f^3$ and $k_f^5$, reflecting contact-interactions without and
with two derivatives. We have investigated the differences in the energy
per particle with respect to the case $\Lambda_0=1.15\,$ GeV where the running 
contact-couplings are zero: $B_3(\Lambda_0) = B_5(\Lambda_0)=0$. For a wide 
range of monopole masses, $0.4\,{\rm GeV} < \Lambda < 2.5\,$GeV, these 
deviations oscillate merely between $+1.5\,$MeV and $-1.0\,$MeV in the whole 
density region $0 < \rho < 0.5\,$fm$^{-3}$. Therefore one can speak of a 
pretty stable nuclear matter saturation curve in our calculation which 
combines pion-exchange dynamics including a regularizing monopole form factor
with (two) scale-dependent short-distance contact-couplings.

\begin{figure}
\begin{center}
\includegraphics[scale=0.5,]{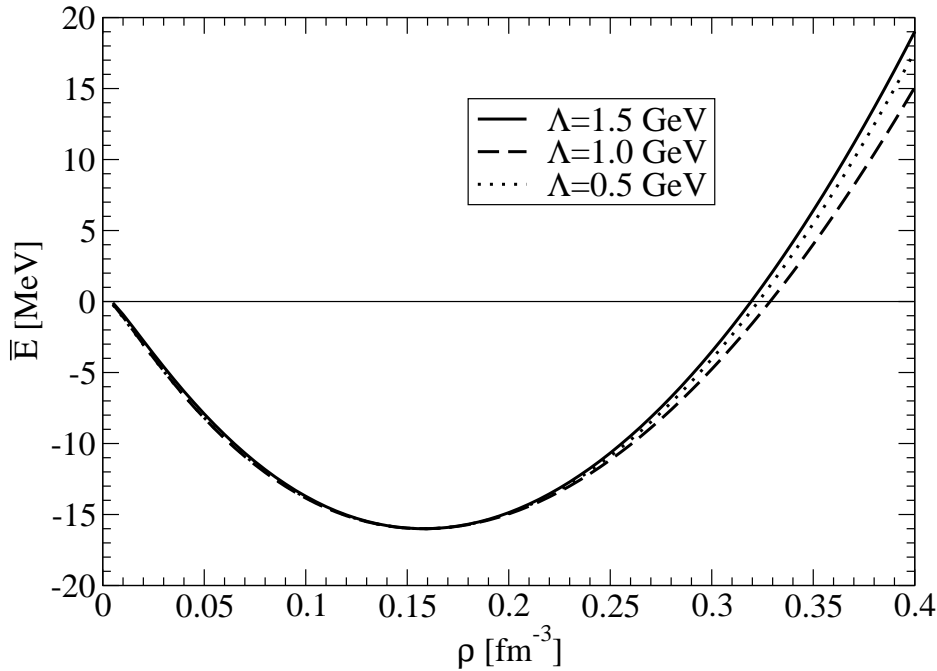}
\end{center}
\vspace{-0.5cm}
\caption{Energy per particle $\bar E(k_f)$ of isospin-symmetric nuclear matter
versus the nucleon density $\rho = 2k_f^3/3\pi^2$. The curves correspond to
three different monopole masses $\Lambda =0.5\,$GeV, $1.0\,$GeV and $1.5\,$GeV 
with the contributions of the compensating contact-couplings $B_{3,5}(\Lambda)$
included.}  
\end{figure}

\section{Chiral  condensate at finite density}
The chiral condensate $\langle 0|\bar q q|0\rangle$ is an order parameter of
spontaneous chiral symmetry breaking. With increasing temperature the chiral
condensate decreases (or ''melts away''). For low temperatures this effect can 
be systematically calculated in chiral perturbation theory.  At three-loop 
order \cite{gerber} the estimate $T_c \simeq 190\,$MeV of the critical 
temperature, where chiral symmetry will be eventually restored, has been 
found. This (extrapolated) value of $T_c$ is remarkably consistent with the
recent result $T_c = (192\pm 8)\,$MeV obtained in numerical simulations of
full QCD on the lattice \cite{cheng}. The chiral condensate drops also with 
increasing baryon density. The leading linear term in the nucleon density 
$\rho$  is readily derived (via the Feynman-Hellmann theorem) by
differentiating the energy density of a nucleonic Fermi gas, $\rho M_N + {\cal
O}(\rho^{5/3})$, with respect to the light quark mass $m_q$. This introduces 
the nucleon sigma-term $\sigma_N =\langle N |m_q \bar q q|N\rangle = m_q \,
\partial M_N/\partial m_q$ as the driving term for the density evolution of 
the chiral condensate. Corrections beyond the linear density approximation 
arise from the $NN$-interactions which transform the nucleonic Fermi gas into 
a nuclear Fermi liquid. Because of the Goldstone boson nature of the pions,
$m_\pi^2\sim m_q$, the explicit pion-exchange dynamics in nuclear matter plays 
a particularly important role for the in-medium chiral condensate  $\langle
\bar q q\rangle(\rho)$. Converting quark mass derivatives into pion mass 
derivative and using the Gell-Mann-Oakes-Renner relation $m_\pi^2 f_\pi^2 = -
m_q \,\langle 0|\bar q q| 0\rangle$ one finds for the ratio of the in-medium
to vacuum chiral condensate:
\begin{equation} {\langle\bar q q\rangle(\rho)\over\langle0|\bar q q|0\rangle} 
= 1 - { \rho \over 2m_\pi f_\pi^2} \bigg[ {2 \sigma_N \over m_\pi} + {d \bar
  E(k_f) \over d m_\pi} \bigg] \,. \end{equation} 
The nucleon sigma-term: 
\begin{equation} \sigma_N = m_q \,{\partial M_N \over\partial m_q} = {m_\pi 
\over   2}\, {\partial M_N \over \partial m_\pi} \,, \end{equation} 
measures that portion of the nucleon mass $M_N$ which arises from the explicit
chiral symmetry breaking in QCD (i.e. the non-vanishing up/down quark mass 
$m_q \ne 0$). The empirical value of $\sigma_N$ as extracted from dispersion
relation analyses of $\pi N$-scattering data is $\sigma_N = (45\pm 8)\,$MeV
\cite{sigma}. 

From our nuclear matter calculation with explicit one- and two-pion exchange 
dynamics we can now easily compute the derivative $d \bar E(k_f)/d m_\pi$ of 
the energy per particle with respect to the pion mass. In addition to the
explicit $m_\pi$-dependence  we consider also the implicit one through the 
nucleon mass: $\partial M_N/\partial m_\pi = 2\sigma_N/m_\pi\simeq 0.67$.
The smaller and less well-known quark mass derivatives of the axial-vector
coupling constant $g_A$ and the pion-decay constant $f_\pi$ are neglected at
present. For example, one-loop chiral perturbation theory gives for the 
pion-decay constant the small derivative: $\partial f_\pi/ \partial m_\pi =
(\bar l_4-1)m_\pi/(8 \pi^2 f_\pi) \simeq 0.063$ \cite{cola}. Furthermore, from
the combination of lattice QCD results for $g_A$ and their chiral
extrapolation \cite{massi} one estimates $m_\pi \,\partial g_A/ \partial m_\pi 
\leq 0.05$ around the physical point. Corrections to $d \bar E(k_f)/d m_\pi$ 
from the $m_\pi$-dependence of the $\pi N$-coupling constant $\sim g_A/
f_\pi$, are therefore expected to be small, less than $5\%$.

\begin{figure}
\begin{center}
\includegraphics[scale=0.45]{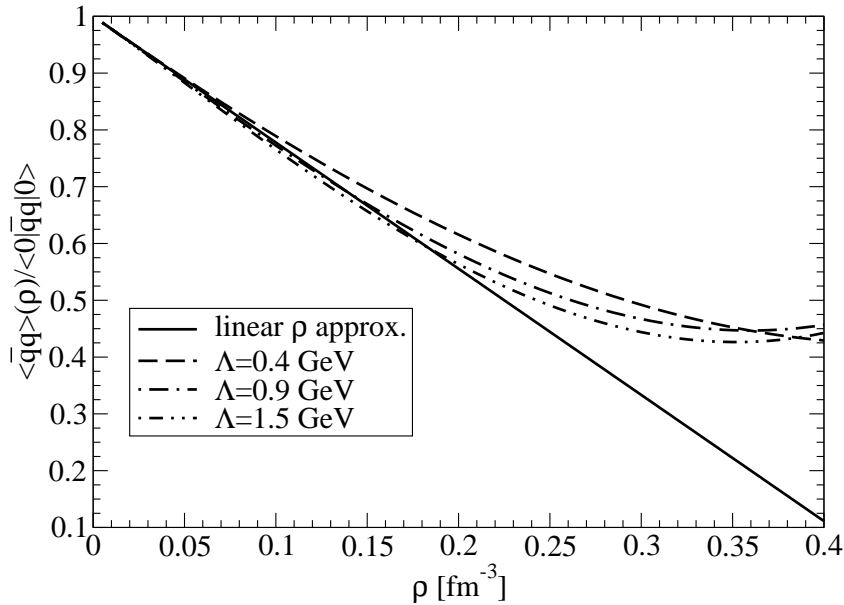}
\end{center}
\vspace{-0.6cm}
\caption{Density dependence of the chiral quark condensate $\langle\bar q q
\rangle(\rho)$.}
\end{figure}

Fig.\,6 shows the dropping chiral condensate $\langle\bar q q \rangle(\rho)$
as a function of the nucleon density up to $\rho = 0.4\,$fm$^{-3}$. The full
line corresponds to the linear density approximation using the central value
$\sigma_N =45\,$MeV. The other curves include the effects of nuclear
correlations through the derivative $d \bar E(k_f)/d m_\pi$ for various values
of the monopole mass $\Lambda$. We have assumed that the short-distance
contact-couplings, $B_3(\Lambda)/M_N^2$ and $B_5(\Lambda)/M_N^4$, which keep 
the saturation curve $\bar E(k_f)$ stable (see section 3) are quark mass
independent. This is a natural assumption, but difficult to substantiate
further. One observes in Fig.\,6  that up to equilibrium density $\rho_0$ the 
corrections beyond the linear density approximation are in fact very small. 
This finding can be taken as an a posteriori justification of the assumption
made in ref.\cite{finelli} about the in-medium behavior of the scalar 
mean-field $\Sigma_S^{(0)}$. At higher densities a trend counteracting chiral
restoration sets in. Similar features have been observed in the earlier chiral 
approach of Lutz et al.\,\cite{lutz} as well as in ref.\cite{rolf} where the
relativistic scalar-vector mean-field phenomenology has been combined with 
estimates of the quark mass derivatives. It is comforting to see that the
explicit one- and two-pion exchange dynamics does not lead to the opposite
trend, namely acceleration of chiral restoration, which would undermine the
foundation of the present approach to nuclear matter, based on the spontaneous
breaking of chiral symmetry (in the vacuum). 

Finally, one should note that the nucleon sigma-term $\sigma_N$ has presently 
a  sizable uncertainty of $\pm 18\%$. Therefore, the error band associated
with the linear density approximation $1-\rho \,\sigma_N/(m_\pi^2 f_\pi^2)$ 
masks practically all effects from nuclear correlations at least up to nuclear
matter saturation density, $\rho_0 = 0.16\,$fm$^{-3}$.

\section{Concluding remarks and outlook}
In this work we have performed a nuclear matter calculation which treats the 
long-range correlations from one- and two-pion exchange explicitly. In
accordance with the recent approach of ref.\cite{achim} based on the $V_{\rm 
low-k}$ potential we find that repulsive three-body terms are essential in 
order to achieve realistic binding and saturation of nuclear matter. As a 
novel feature we have introduced a pion-nucleon monopole form factor as a
regulator to eliminate high-momentum components from the interactions. The 
dependence on  the ''resolution'' parameter, the monopole mass $\Lambda$, can 
be perfectly counterbalanced by two running short-distance contact-couplings 
$B_{3,5}(\Lambda)$. The resulting nuclear matter equation of state $\bar
E(k_f)$ is stable against variations of $\Lambda$. We have taken the pion mass 
derivative $d\bar E(k_f)/d m_\pi$ and obtained small corrections to the 
in-medium chiral condensate  $\langle\bar q q\rangle(\rho)$ beyond the linear 
density approximation. 

A more detailed comparison of the present nuclear matter calculation with the 
one based on the $V_{\rm low-k}$ potential \cite{achim} offers prospects for a 
better understanding of the relevant short-range $NN$-dynamics, which so far 
enters in the form of two adjusted contact-couplings $B_{3,5}(\Lambda)$. Work 
along these lines is in progress \cite{prog}.   
\section*{Acknowledgment}
We thank Achim Schwenk for many useful discussions.   

\end{document}